\documentstyle[12pt]{article}
\begin{document}
\thispagestyle{empty}
\begin{center}
\LARGE \tt \bf{Cosmological density perturbation on spin dominated universes and COBE data}
\end{center}
\vspace{2.5cm}
\begin{center}{\large L.C. Garcia de Andrade\footnote{Departamento de F\'{\i}sica Teorica - UERJ.
Rua S\~{a}o Fco. Xavier 524, Rio de Janeiro, RJ
Maracan\~{a}, CEP:20550-003 , Brasil.
{e-mail: garcia@dft.if.uerj.br}}}
\end{center}
\vspace{2.0cm}
\begin{abstract}
The evolution equation of cosmic density perturbation of a spin dominated universe in Einstein-Cartan gravity is obtained.Examples of inflationary cosmologies with spin are also given where we compute the density perturbations in terms of redshift and show that spin-torsion effects produce a growth on the inhomogeneity of the universe as the redshift is high.Spin-torsion effects are redshifted out when the Universe expands and would be rarely of importance in present day phase of the universe.Comparison with the primordial density fluctuation of COBE data is also obtained.From our computations a COBE type experiment could be proposed to measure the spin-torsion effects in large scale structure objects such as spiral Galaxies and globular clusters.  
\end{abstract}
\vspace{2.0cm}
\newpage
\section{Introduction}
The study of in the formation structure in the Universe \cite{1,2} has given a renewal interest 
with the advent and success of extraterrestrial experiments like the Cosmic Background Explorer (COBE) and its data \cite{3}.Recently the investigation of the density perturbations in alternative theories of gravity such as Brans-Dicke due to the presence of scalar inflaton or dilatonic fields in inflationary scenario have been undertaken by Fabris and his group \cite{4,5,6}.Besides computation of density perturbations in global textures in the realm of Newtonian cosmology have been investigated by  Ribeiro and Letelier \cite{7}.Earlier Palle \cite{8} and myself \cite{9} have independently investigate the idea of primordial density flucuations in Einstein-Cartan gravity and show that Einstein-Cartan cosmology is compatiible with COBE data.Nevertheless Palle has not continued the investigation of the primordial fluctuations providing specific examples that could be matched with future experimental data base of future experiments such as the Boomerang and other experimental devices placed on the Earth orbit.In this paper we propose to fill this gap.The paper is organized as follows:In section 2 we give a simple derivation of the evolution of density perturbations and its dynamics.In section 3 we give twoo simple examples,the first is of a two fluid Friedmann Universes embbeded on each other where one does not contain spinning particles and the other can be analyzed using Einstein-Cartan gravity and the other is a two fluid this time not closed universes as before but spatially flat and possesing dilatonic fields.In both cases the density perturbations are expressed in terms of the redshift parameter z and we show that spin-torsion density contribution is stronger when the redshifts are higher.Data from elliptical Galaxies and globular cluster are also used to provide bounds for the density perturbations and spin contributions to them.In section 4 gravitational stability of the Friedmann metric in Einstein-Cartan gravity is also discussed since this is a important topic in Galaxy formation.Besides a lower boud to the spin-torsion fluctuation is found from this analysis and COBE data.  
\section{Evolution Equation of Density Perturbation}
Investigation of linear density fluctuations \cite{1,2} in the context of general relativistic cosmology have proved to be useful in the study of structure formation like galaxy formation for example.Later on we developed \cite{9} some of his ideas to solutions of the Einstein-Cartan cosmology with inflatons and dilatons.In this section we show that
the evolution of density fluctuations can be derived from the Einstein-Cartan field equations and the associated conservation equation for the matter and spin-torsion density in analogous way that it is done in General Relativity as long as some simple assumptions are made on the galatic spinning fluid.A stationary metric here is not needed since we are not considering that the spin of the intrinsic particles affect appreciably the rotation of galaxies.Therefore the usual Friedmann metric to investigate linear perturbations in Einstein-Cartan gravity.Let us begin by considering the Friedmann metric
\begin{equation}
ds^{2}=dt^{2}-a^{2}(dx^{2}+dy^{2}+dz^{2})
\label{1}
\end{equation}   
where $a(t)$ is the cosmic scale.The Einstein-Cartan equations are given by 
\begin{equation}
H^{2}=\frac{8{\pi}G}{3}({\rho}_{eff}-2{\pi}G{\sigma}^{2})
\label{2}
\end{equation}
and 
\begin{equation}
H^{2}+\dot{H}=-\frac{4{\pi}G}{3}({\rho}_{eff}-8{\pi}G{\sigma}^{2})
\label{3}
\end{equation}
where $H=\frac{\dot{a}}{a}$ is the Hubble parameter in terms of time.Where in short we use the following notations
\begin{equation}
{\rho}_{eff}={\dot{\phi}}^{2}+V
\label{4}
\end{equation}
and
\begin{equation}
p_{eff}={\dot{\phi}}^{2}-V
\label{5}
\end{equation}
where ${\phi}$ is the inflaton field ,$V$ is the inflaton potential and ${\sigma}^{2}=<S_{ij}S^{ij}>$ is the averaged squared of the spin density tensor $S_{ij}$.  
By making use of the definition ${\rho}={\rho}_{eff}-2{\pi}G{\sigma}^{2}$ and $p=p_{eff}-2{\pi}G{\sigma}^{2}$ allow us to write the conservation equation in Einstein-Cartan Gravity \cite{10} as
\begin{equation}
\frac{d}{dR}({{\rho}R^{3}})=-3pR^{2}
\label{6}
\end{equation}
These substitutions allow us to reduce our problem of computing density perturbations formally similar to the general relativistic one.The necessary assumption to make our task easier is to consider the pressure $p$ vanishes 
which yields
\begin{equation}
p_{eff}=2{\pi}G{\sigma}^{2}
\label{7}
\end{equation}
With these assumptions equation (\ref{6}) reduces to
\begin{equation}
\frac{d}{dt}({{\rho}R^{3}})=0
\label{8}
\end{equation}
where now we are ready to apply the traditional perturbation method on the
cosmological density perturbation by making use of the following definitions
\begin{equation}
\delta=\frac{{\rho}-{\rho}_{b}}{{\rho}_{b}}
\label{9}
\end{equation}
and
\begin{equation}
a=R(t)+{\delta}R(t)
\label{10}
\end{equation}
and
\begin{equation}
{\sigma}^{2}={\sigma}^{2}_{b}(1+\frac{{\delta}{\sigma}^{2}}{{\sigma}^{2}})
\label{11}
\end{equation}
Here the index b denotes background quantities.From the equation (\ref{8}) one obtains
\begin{equation}
{\delta}=-3\frac{{\delta}R}{R}
\label{12}
\end{equation}
following the lines and procedures of General Relativity we find the following evolution equation
yields
\begin{equation}
{\ddot{\delta}}+2{H_{0}}^{2}{\dot{\delta}}+10{\pi}^{2}G^{2}{{\sigma}_{b}}^{2}{\delta}=0
\label{13}
\end{equation}
To solve this differential equation we need to compute the background spin-torsion density.This can be easily accomplished if one equates equations (\ref{2}) and (\ref{3}) and substitute (\ref{7}) to obtain 
\begin{equation}
{\rho}_{eff}=\frac{9}{2}{\pi}G{{\sigma}_{b}}^{2}
\label{14}
\end{equation}
Substitution of this last expression into the conservation equation for the effective density
yields
\begin{equation}
{\dot{{{\sigma}_{b}}^{2}}}+3H_{0}{{\sigma}_{b}}^{2}=0
\label{15}
\end{equation}
which solution is
\begin{equation}
{{\sigma}_{b}}^{2}=e^{-3H_{0}t}
\label{16}
\end{equation}
to simplify matters without loosing too much physical insight one could assume that the spin-torsion background can be expanded in the form
\begin{equation}
{{\sigma}_{b}}^{2}={(1+3H_{0}t)}^{-1}={(3H_{0}t)}^{-1}
\label{17}
\end{equation}
where we have made the hypothesis $H_{0}t>>\frac{1}{3}$.Substitution of this result into expression (\ref{13}) yields the final form of the evolution equation ready now to be solved 
\begin{equation}
{\ddot{\delta}}+2{\alpha}{\dot{\delta}}+{\gamma}\frac{{\delta}}{t}=0
\label{18}
\end{equation}
where ${\alpha}=2{H_{0}}^{2}$ and ${\gamma}=\frac{-10{\pi}G}{3H_{0}}$.To deduce the evolution equation (\ref{18}) we assumed that the perturbed metric is the de Sitter metric where 
$H_{0}$ is a constant.To solve this evolution equation let us assume as usual a solution of the type ${\delta}=t^{m}$ where $m$ is a real constant to be determined from an algebraic equation.Substitution of this hint into the differential  equation (\ref{18}) yields the following algebraic equation
\begin{equation}
m^{2}+(1+{\alpha})m+{\gamma}=0
\label{19}
\end{equation}
the solution of this last equation left us with two solutions of the type ${\delta}_{+}=t^{m_{+}}$ and $ {\delta}_{-}=t^{m_{-}}$ where $m_{+}=\frac{5{\pi}G}{3H_{0}}$ and $m_{-}=-(1+{H_{0}}^{2})$ are the respective elementary solutions of the algebraic equation (\ref{19}). 
The spinning fluid of the Early  Universe consists here of protons and neutrons which are fermions producing the torsion of spacetime.The fluid this time is not a dilatonic fluid and there is no potential or inflaton.The resulting evolution euqtion is simply the general relativistic equation
\begin{equation}
{\ddot{\delta}}+2\frac{\dot{R}}{R}{\dot{\delta}}-4{\pi}^{2}G{{\rho}_{b}}_{eff}{\delta}=0
\label{20}
\end{equation}
where now ${{\rho}_{b}}_{eff}={\rho}_{b}-2{\pi}G{{\sigma}_{b}}^{2}$.To be able to express the evolution equation in terms of the cosmic time it is enough to remenber taht the mass density of the background is given by
${\rho}_{b}=\frac{M}{R^{3}}$ and the spin-torsion density can be obtained remenbering that the spin-torsion density tensor can be written as $|S_{ij}|=S=\frac{nh}{R^{3}}$ where n is the number of fermions and h is the Planck constant.The Friedmann equation thus becomes
\begin{equation}
{\dot{R}}^{2}=\frac{8{\pi}G}{3}(\frac{M}{R}-\frac{2{\pi}GS^{2}}{R^{4}})
\label{21}
\end{equation}
allow us to obtain explicitly the dependence of the cosmic scale R with time.At this point since we are mainly interested in the extremum case of the role played by torsion and spin on Galaxy formation we assume here that the first term may be dropped,(as could happens inside black holes for example) which reduces the above equation to
\begin{equation}
{\dot{R}}^{2}=-\frac{16{\pi}^{2}G^{2}S^{2}}{3R^{6}}
\label{22}
\end{equation}
Substitution of $R=t^{p}$ where p is a real number into the equation (\ref{22}) one obtains the following algebraic equation 
\begin{equation}
4p^{2}-4p+1=0
\label{23}
\end{equation}
Solution of this equation yields $R(t)=t^{\frac{1}{2}}$.Substitution of this result into formula (\ref{22}) yields
\begin{equation}
{\ddot{\delta}}+2\frac{\dot{\delta}}{t}+16{\pi}^{2}G^{2}\frac{n^{2}h^{2}}{3t^{3}}{\delta}=0
\label{24}
\end{equation}
Substitution of the ansatz ${\delta}=t^{m}$ into equation (\ref{28}) yields
\begin{equation}
m^{2}+3m-4{\pi}GM=0
\label{25}
\end{equation}
which yields the following solutions ${\delta}_{+}=t^{3}$ and 
${\delta}_{-}=1$ which possess the interesting physical feature that the second equation coincides with the general relativistic solution while the first solution seems to be a carachteristic of the Einstein-Cartan Cosmology.Our result also shows that the gravitational instability is enhanced by torsion instead of holding it.This result have already been expected by the analysis done by Hehl et al. \cite{10} many years ago.
\section{Density perturbation in Spin cosmology}
When D.Palle \cite{8} computed the evolution equation of density perturbation for a Bianchi type III model in Einstein-Cartan cosmology.It is well know in standard general relativistic cosmology that specially in the particular case of Newtonian cosmology simple models can be designed to obtain the density perturbations without solving the evolution equation.In this note we propose a two fluid very simple model in Einstein-Cartan gravity to compute the cosmological density perturbations without solving the evolution equation of density perturbations.In the absence of spin our results reduce to the ones in General Relativity \cite{2}.Dominance of the spin-torsion density in the early epochs of the Universe occurs in the case of density perturbations for the matter phase while at the present epoch spin-torsion stronger contribution is to the density perturbations is on radiation era.This fact seems to be physically explainable since the early epochs of the Universe are carachterize by the fact that the matter-radiation coupling is much stronger than in the present day Universe.Let us start by considering the Friedmann metric 
\begin{equation}
ds^{2}=dt^{2}-a^{2}(t)(\frac{dr^{2}}{1-kr^{2}}+r^{2}(d{\theta}^{2}+sin^{2}{\theta}d{\phi}^{2}))
\label{26}
\end{equation}
where $a(t)$ is connected to the Hubble parameter H by the relation $H=\frac{\dot{a}}{a}$ and k is the spatial curvature constant which is $k=-1,0,+1$ according the 3-space is open,flat or closed respectively.Considering a spherical region of radius ${\lambda}>d_{H}$ which is the Hubble radius containing spinning matter with mean density ${\rho}_{1}$,embedded in a $k=0$ Friedmann universe ${\rho}_{0}$ where ${\rho}_{1}={\rho}_{0}+{\delta}{\rho}$, ${\delta}{\rho}>0$ and represents a small density perturbation.As point it out by Padmanabhan \cite{5} the spherical symmetry implies that the inner region is not affected by the matter outside.Thus the inner region evolves as a $k=+1$ Friedmann Universe .The two regions form a kind of two fluid the inner fluid being a spinning fluid obtained from the perturbation.Since the inner fluid is a spin fluid it may obey the Einstein-Cartan gravity with the following form
\begin{equation}
{H_{1}}^{2}+\frac{1}{{a^{2}}_{1}}=\frac{8{\pi}G}{3}({\rho}_{1}-2{\pi}G{{\sigma}^{2}})
\label{27}
\end{equation}
where ${\sigma}^{2}$ represents the averaged squared value of the spin-torsion tensor.The outer fluid obey the GR equation
\begin{equation}
{H_{0}}^{2}=\frac{8{\pi}G}{3}({\rho}_{0})
\label{28}
\end{equation}
where both universes are compared, the perturbed and the background universe when their expansion rates are equal are equal,or we compare their densities at the time t when $H_{1}=H_{0}$.Since
\begin{equation}
\frac{{\delta}{\rho}}{{\rho}_{0}}=\frac{{\rho}_{1}-{\rho}_{0}}{{\rho}_{0}}
\label{29}
\end{equation}
Performing the above substitution we obtain
\begin{equation}
\frac{{\delta}{\rho}}{{\rho}_{0}}=\frac{1}{{8{\pi}G}{a^{2}}_{1}{\rho}_{0}}+\frac{2{\pi}G{\sigma}^{2}}{{\rho}_{0}}
\label{30}
\end{equation}
To compute this equation in terms of the cosmic expansion $a$ we simply remember that the spin-torsion density is proportional to $\frac{{S_{0}}^{2}}{a^{6}}$.where ${S_{0}}^{2}=n^{2}h^{2}$ is the spin tensor squared and n is equal to the number of nucleons in the Universe and h is the Planck constant.Substituting this value into the equation (\ref{30}) and considering that $a_{1}$ is approximatly equal to $a_{0}$ allows us to see how the $\frac{{\delta}{\rho}}{{\rho}_{0}}$ scales with $a$.Since in the radiation dominated era ${\rho}_{0}{\alpha}a^{-4}$ and ${\rho}_{0}{\alpha}a^{-3}$ in the matter dominated phase, substitution of these  values into the equation (\ref{30}) yields respectively
\begin{equation}
\frac{{\delta}{\rho}}{{\rho}_{0}}|_{R}=\frac{a^{2}}{8{\pi}G}+{\beta}G{S_{0}}^{2}{a}^{-2}
\label{31}
\end{equation}
and
\begin{equation}
\frac{{\delta}{\rho}}{{\rho}_{0}}|_{M}=\frac{a}{8{\pi}G}+{\beta}G{S^{2}}_{0}{a}^{-3}
\label{32}
\end{equation}
others zero.From the dependence of the spin-torsion term on the coosmic factor $a$ we already note that the spin-torsion contributions would be redshifted with expansion.Here ${\beta}$ is a constant.To write down these expressions in terms of cosmic time to be able to check the observational results with our computations we need to solve write {\ref{36}) in terms of the cosmic scale factor a in terms of time.Solution of this equation in terms of time yields the following results $a|_{R}{\alpha}t^{\frac{1}{2}}$ and $a|_{M}{\alpha}t^{\frac{2}{3}}$.Substitution of these values into the expressions (\ref{31}) and (\ref{32}) we obtain respectively
\begin{equation}
\frac{{\delta}{\rho}}{{\rho}_{0}}|_{R}=\frac{t}{8{\pi}G}+{\beta}G{S_{0}}^{2}{t}^{-1}
\label{33}
\end{equation}
and
\begin{equation}
\frac{{\delta}{\rho}}{{\rho}_{0}}|_{M}=\frac{t^{\frac{2}{3}}}{8{\pi}G}+{\beta}G{S_{0}}^{2}{t}^{-2}
\label{34}
\end{equation}
Notice therefore that in the case of early epochs in the Universe where $t-0$ the spin-torsion effects appear to be stronger on matter perturbation contributing to the decoupling of matter and radiation this is explained since the torsion acts only on fermions and do not clearly interacts with radiation.Nevertheless in the present epoch the spin-torsion decays very quick which is the reason the detection of torsion is difficult froom the cosmological point of  view nowadays.To have a still more simple vision of the problem we express the formula for the density perturbation of the matter in terms of the redshift z.This can be done remenber the horizon problem and using the formula
\begin{equation}
a_{H}=a(t)\int{\frac{dt'}{a(t')}}
\label{35}
\end{equation}
and taking the value $a{\alpha}t^{\frac{2}{3}}$ obtained previously for the matter case and substituting it to expression (\ref{35}) and performing the integration one obtains $a_{H}=3t$.In turn substitution of this result into the expression for the age of the Universe \cite{1} one obtains
\begin{equation}
t=\frac{1}{3}a_{H}=\frac{2H_{0}^{-1}}{(1+z)^{\frac{3}{2}}}
\label{36}
\end{equation}
and finally substitution of this value into equation (\ref{34})
\begin{equation}
{\delta}(z)|_{M}=\frac{{\delta}{\rho}}{{\rho}_{0}}|_{M}=\frac{(2H_{0}^{-1})^{\frac{2}{3}}}{8{\pi}G(1+z)}+\frac{H_{0}^{3}{\beta}G{S_{0}}^{2}(1+z)^{3}}{8}
\label{37}
\end{equation}
From this last expresion is easy to check that spin-torsion effects are dominant for high redshifts and since the higher redshift objets are far away from us one may realize that spin-torsion effects are redshifted when the Universe expands for example in the case of inflationary cosmology.Specially strong effects can be obtained at the decoupling time where $z=10^{3}$.At this point let us consider the other example of flat spatially cosmological models embedded in each other and obeying the following field equations
\begin{equation}
{H_{1}}^{2}=\frac{8{\pi}G}{3}({\rho}_{1}+\frac{1}{2}{\dot{\phi}}^{2}-2{\pi}G{{\sigma}^{2}})
\label{38}
\end{equation} 
and
\begin{equation}
{H_{0}}^{2}=\frac{8{\pi}G}{3}({\rho}_{0})
\label{39}
\end{equation}
After a simple algebra one obtains the expression for the density perturbation
\begin{equation}
\frac{{\delta}{\rho}}{{\rho}_{0}}=\frac{2{\pi}G{\sigma}^{2}}{{\rho}_{0}}-\frac{1}{2}\frac{{\dot{\phi}}^{2}}{{\rho}_{0}}
\label{40}
\end{equation}
Since from the already known solution of de Sitter inflationary cosmology with spin-torsion density we have \cite{8}
\begin{equation}
{\dot{\phi}}^{2}={\pi}G{\sigma}^{2}
\label{41}
\end{equation}
Substitution of this last expression into the density perturbation reduces this formula to
\begin{equation}
\frac{{\delta}{\rho}}{{\rho}_{0}}=\frac{\frac{3}{2}{\pi}G{\sigma}^{2}}{{\rho}_{0}}
\label{42}
\end{equation}
Now taking into account the behaviour of matter phase ${\rho}_{0}{\alpha}a^{-3}$ and the relation between the redshift and the cosmic parameter $a$ as $a=(1+z)^{-1}$ one obtains
\begin{equation}
\frac{{\delta}{\rho}}{{\rho}_{0}}=\frac{\frac{3}{2}{\pi}G{\sigma}^{2}}{(1+z)^{3}}
\label{43}
\end{equation}
This expression allow us to give numerical estimates to the density perturbations as far as spin-torsion effects are concerned.Since from de Sabbata and Sivaram \cite{11} we know that at the Planck era we have $\frac{G{\sigma}^{2}}{c^{4}}=10^{87}$ and that at the Planck era $t_{Pl}=10^{-43}s$ the expression 
\begin{equation}
\frac{{\delta}{\rho}}{{\rho}_{0}}=\frac{3}{2}{\pi}G{{\sigma}_{Pl}}^{2}{t_{Pl}}^{2}=10>1
\label{44}
\end{equation}
and the linear perturbation density approximation would break down and nonlinear perturbation theory would have to be used.On the other hand making use of the spin-torsion critical value $\frac{G{\sigma}^{2}}{c^{4}}=10^{-30}$ and the value of the redshift at the decoupling between matter and radiation $z=10^{3}$ one obtains the value of
\begin{equation}
\frac{{\delta}{\rho}}{{\rho}_{0}}=\frac{\frac{3}{2}{\pi}G{\sigma}^{2}}{(1+z)^{3}}=10^{-33}
\label{45}
\end{equation}
which is an extremely low value for being detected by the COBE experimental devices.Nevertheless an estimate within the COBE capabilities may be obtained by making use of two Astronomical data, namely the spiral Galaxies and cluster energy densities of ${\rho}_{0}=10^{-25}g.cm^{-3}$ and ${\rho}_{0}=10^{-28} g.cm^{-3}$ respectively which yield for the spin-torsion contribution values of
\begin{equation}
\frac{{\delta}{\rho}}{{\rho}_{0}}=10^{-5}
\label{46}
\end{equation}
and
\begin{equation}
\frac{{\delta}{\rho}}{{\rho}_{0}}=10^{-2}
\label{47}
\end{equation}
well within the COBE data measurement capabilities.Here we also used the above critical spin-torsion density value.
\section{Gravitational Stability of Friedmann Metric in Spin Cosmology}
Now we turn to the problem of the stability of the Friedmann metric in Einstein-Cartan cosmology.Nurgaliev and Ponomariev \cite{12} investigated the early evolutionary stages of the Universe in the realm of Einstein-Cartan cosmology by considering an ideal fluid of nonpolarised fermions.They show that for this particular case the Friedmann solution was stable to small homogeneous and isotropic perturbations.They also conclude that the increase in entropy could lead to the evolution of the initially small stable oscillations into large ones.The Universe would begin after a definite number of oscillations.Calculations of small perturbations in their model would shown some instability which would prepare for the necessary initial conditions for the growth of pertubations in the nonlinear regime.Now we make use of their result to olace a lower limit to the spin-torsion primordial density fluctuation obtained from the Einstein-Cartan gravity and COBE satellite data \cite{1,2}.In this way structure formation like Galaxies formation would not suffer a great influence from torsion \cite{11}.It is also shown that at early stages of the Universe Bianchi type III models with expansion and rotation may not depend at all from torsion.Other Bianchi types like an oscillating Bianchi type IX model in Einstein-Cartan \cite{13} gravity have also been recently investigated.The Friedmann equation thus becomes
\begin{equation}
\frac{{\ddot{a}}}{{a}}=\frac{4{\pi}G}{3}({\rho}-8{\pi}G{\sigma}^{2})
\label{48}
\end{equation}
Making an homogeneous and isotropic small perturbation on the Friedmann yields
\begin{equation}
\frac{{\delta}{\ddot{a}}}{{\delta}{a}}=-\frac{4{\pi}G}{3}({\rho}-8{\pi}G{\sigma}^{2})-\frac{4{\pi}G}{3}(\frac{{\delta}{\rho}}{{\delta}a}-8{\pi}G\frac{{\delta}{\sigma}^{2}}{{\delta}{a}})
\label{49}
\end{equation}
Substitution of the well-known relation
\begin{equation}
\frac{{\delta}{\rho}}{{\delta}a}=-3\frac{\rho}{a}
\label{50}
\end{equation}
we obtain
\begin{equation}
\frac{{\delta}{\ddot{a}}}{{\delta}{a}}=-\frac{4{\pi}G}{3}({\rho}-8{\pi}G{\sigma}^{2})+\frac{4{\pi}G}{3}(\frac{{\rho}}{a}+8{\pi}G\frac{{\delta}{\sigma}^{2}}{{\delta}{a}})
\label{51}
\end{equation}
and the stability condition $\frac{{\delta}{\ddot{a}}}{{\delta}{a}}<0$ implies 
\begin{equation}
(1-8{\pi}G\frac{{\delta}{\sigma}^{2}}{{\delta}{\rho}})<0
\label{52}  
\end{equation}
Since the matter density ${\rho}>0$ this implies the following condition
\begin{equation}
{{\delta}{\sigma}^{2}}>\frac{1}{8{\pi}G}\frac{{\delta}{\rho}}{{\rho}_{0}}{\rho}_{0}
\label{53}  
\end{equation}
where we take${\rho}_{0}=10^{-31}g{cm}^{-3}$ as the matter density of the Universe and from the COBE data $\frac{{\delta}{\rho}}{{\rho}_{0}}=10^{-5}$.From these data formula (\ref{53}) yields the following lower limit for the spin-torsion fluctuation as
\begin{equation}
{\delta}{\sigma}^{2}>10^{-28} cgs units
\label{54}
\end{equation}
this result was expected since the spin-torsion density decreases with the expansion and is redshifted with inflation.This conjecture has been proposed recently by Ramos and myself \cite{14}.As pointed out by Nurgaliev and Ponomariev \cite{12} the increase in the entropy may trigger the growth in the inhomogeneities.There is no compelling reason to believe that this would not happen here.Moreover Nurgaliev and Piskareva \cite{15} have also investigate the structural stability of cosmological models in Einstein-Cartan gravity.A more detailed investigation of the matters discussed here including the Bianchi type IX oscilating solution in Einstein-Cartan cosmology may appear elsewhere.
\section*{Acknowledgement}
I am very much indebt to Professors P.S.Letelier,I.Shapiro and my colleague Rudnei de Oliveira for helpful discussions on the subject of this paper.Thanks are also due to an unknown referee for useful comments.Special thanks go to Dr.Andre Ribeiro for helping me with Astronomical data.Financial support from CNPq. and UERJ is gratefully acknowledged.

\end{document}